\documentclass[aps,pra,twocolumn,a4paper]{revtex4-1}

\usepackage{mathptmx}
\usepackage{graphicx}
\usepackage[utf8]{inputenc}
\usepackage{textcomp}
\usepackage{braket}
\usepackage{bbm}
\usepackage[hidelinks,colorlinks]{hyperref}
\hypersetup{urlcolor=[rgb]{0,0,0.5},citecolor=[rgb]{0,0.5,0},linkcolor=[rgb]{0.5,0,0}}
\usepackage[all]{hypcap}

\begin{document}

\title{An integrated quantum repeater at telecom wavelength with single atoms in optical fiber cavities}
\author{Manuel Uphoff}
\author{Manuel Brekenfeld}
\author{Gerhard Rempe}
\author{Stephan Ritter}
\email{stephan.ritter@mpq.mpg.de}
\affiliation{Max-Planck-Institut f\"ur Quantenoptik, Hans-Kopfermann-Str. 1, 85748 Garching, Germany}

\begin{abstract}
Quantum repeaters promise to enable quantum networks over global distances by circumventing the exponential decrease in success probability inherent in direct photon transmission. We propose a realistic, functionally integrated quantum-repeater implementation based on single atoms in optical cavities. Entanglement is directly generated between the single-atom quantum memory and a photon at telecom wavelength. The latter is collected with high efficiency and adjustable temporal and spectral properties into a spatially well-defined cavity mode. It is heralded by a near-infrared photon emitted from a second, orthogonal cavity. Entanglement between two remote quantum memories can be generated via an optical Bell-state measurement, while we propose entanglement swapping based on a highly efficient, cavity-assisted atom-atom gate. Our quantum-repeater scheme eliminates any requirement for wavelength conversion such that only a single system is needed at each node. We investigate a particular implementation with rubidium and realistic parameters for Fabry--Perot cavities based on CO$_2$ laser-machined optical fibers. We show that the scheme enables the implementation of a rather simple quantum repeater that outperforms direct entanglement generation over large distances and does not require any improvements in technology beyond the state of the art.
\end{abstract}

\maketitle

\begin{figure*}
\includegraphics[width=2.0\columnwidth]{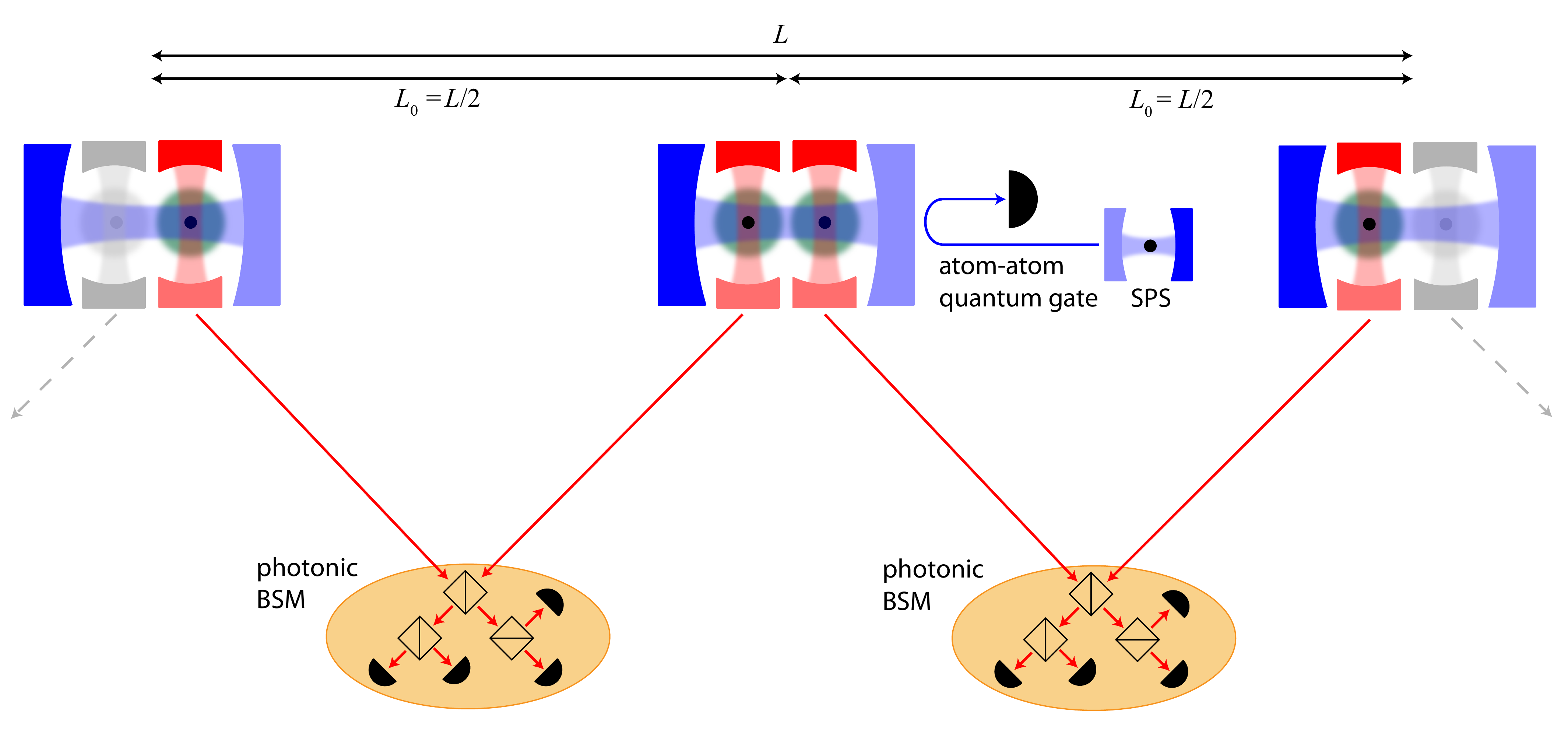}
\caption{Basic quantum repeater scheme featuring single atoms in optical cavities and telecom photons. A single repeater node consists of a heralding cavity (blue, horizontal cavity mode axis) and two telecom-wavelength entangling cavities (red, vertical mode axis). The atoms (black dots) are individually controlled by laser beams perpendicular to the image plane (indicated by green circles around the atoms). Two nodes separated by distance $L/2$ are entangled by first creating atom-photon entanglement locally at each of the nodes and then performing an optical Bell-state measurement (BSM) at a distance $L/4$ from each node. Entanglement swapping between pairs of nodes is implemented at the central node. A cavity-assisted quantum gate is performed on the two atoms via reflection of a single photon originating from a cavity-based single-photon source (SPS). Subsequent detection of the atomic quantum states in suitable bases allows for an unambiguous determination of the two-particle Bell state. This results in an entangled state between the two outermost nodes separated by a distance $L$. This basic unit can be extended to a larger number of swap levels by appending further quantum repeater nodes at the ends of the quantum repeater link (cf. the additional entangling cavities in gray).}
\label{fig:Repeater}
\end{figure*}

\section{Introduction}
Heralded entanglement between particles separated by large distances is a valuable resource in quantum communication. The applications range from fundamental tests of quantum physics like loophole-free Bell tests \cite{Brunner2014,Hensen2015} to device-independent quantum key distribution \cite{Acin2006,Vazirani2014}. The distribution of entanglement via the direct transmission of photons via fiber optics is practically impossible for large distances, because the inevitable losses in optical fibers cause an exponential decrease of the success rate with distance. This can be overcome by dividing the distance into smaller segments with quantum repeater nodes in between \cite{Briegel1998}.

The performance of quantum repeaters can be characterized by the rate at which pairs of quantum memories separated by a given distance can be entangled with high fidelity. The obvious benchmark for a quantum repeater to beat is the rate achievable via direct transmission. The critical parameter in the latter case is the attenuation length of the optical fiber, which is maximal at telecom wavelengths around 1.3\,\textmu m and 1.5\,\textmu m, where the absorption in optical fibers is low \cite{Hubel2007,Inagaki2013}. In order to keep the technological overhead minimal, it is therefore essential to operate a quantum repeater at a telecom wavelength. Additional requirements for an efficient quantum repeater are quantum memories with coherence times by far exceeding the average time required for the protocol to succeed and high-efficiency and high-fidelity implementations of all subparts of the protocol. These include entanglement generation and entanglement distribution via photons as well as entanglement swapping using single-qubit operations, two-qubit operations, and state readout. Against this backdrop, various systems have been proposed for the implementation of a quantum repeater, e.g., atomic ensembles \cite{Duan2001}, single neutral atoms and ions \cite{Childress2005}, nitrogen vacancy centers \cite{Childress2006}, quantum dots \cite{Childress2005}, and ion-doped solids \cite{Simon2007}.

Single atoms in optical cavities are especially promising \cite{Borregaard2015}, because they can be isolated from the environment to provide long coherence times \cite{Treutlein2004} and have been shown to be an efficient light-matter interface \cite{Ritter2012}. Several subparts of a quantum-repeater protocol have been successfully implemented with these systems at near-infrared wavelengths, e.g., the generation of atom-photon entanglement \cite{Wilk2007a}, an atom-photon quantum gate \cite{Reiserer2014}, and the heralded storage of a photonic quantum bit \cite{Kalb2015}. Demonstrations of a quantum repeater that go beyond proof of concept will have to find a way to combine these operations with telecom-wavelength photons. This has recently sparked very active research in external devices that convert photonic qubits at wavelengths in the near-infrared to telecom wavelengths \cite{Radnaev2010,Albrecht2014}. An alternative route being pursued is the generation of an entangled photon pair via spontaneous parametric down-conversion, with one photon at telecom wavelength and the other in the near-infrared, followed by storage of the near-infrared photon in a quantum memory \cite{Saglamyurek2011,Clausen2011}. While these strategies seem straightforward, they come at the price of a large technological overhead and reduced efficiency.

 Here, we therefore propose to perform the required operations directly at telecom wavelengths, thereby avoiding any issues that arise from combining the different technologies. We describe a realistic scheme for a simple quantum repeater that can be implemented using current technology and show that it is capable of outperforming schemes based on direct transmission. The basic unit is illustrated in Fig.\,\ref{fig:Repeater}. Heralded entanglement between a single atom in a crossed-cavity setup and a photon at telecom wavelength is created. The telecom photons from two remote atoms are sent to a photonic Bell-state analyzer to create heralded remote entanglement. Once the two atoms at the central node are each entangled with a different remote node, an atomic Bell-state measurement (BSM) is performed for entanglement swapping. The scheme can be implemented with current technology and provides a clear path to an experimental demonstration of a quantum repeater.

In the following, we describe the implementation of the individual parts of the scheme. In Sec.\,\ref{sec:EntanglementScheme} we propose a way to directly generate entanglement between single atoms and a telecom-wavelength photon using a cascaded scheme. We will estimate the performance of a particular implementation using $^{87}$Rb and realistic cavity parameters. In Sec.\,\ref{sec:RemoteEntanglement} we investigate the indistinguishability of the telecom photons created in this way and show that they are well suited for a photonic BSM that entangles the two remote single-atom quantum memories. Entanglement swapping by atomic BSM will be the topic of Sec.\,\ref{sec:EntanglementSwapping}. In Sec.\,\ref{sec:Repeater} we analyze the performance of the full repeater scheme and show that it can outperform entanglement generation based on direct transmission. We furthermore discuss the prospects of integrating entanglement purification in our scheme. While all required components can be implemented in our cavity-based approach, we show that under the assumption of realistic fidelities, entanglement purification is of little use for a quantum repeater over moderate distances with a small number of swap levels.

\begin{figure*}
\includegraphics[width=2.0\columnwidth]{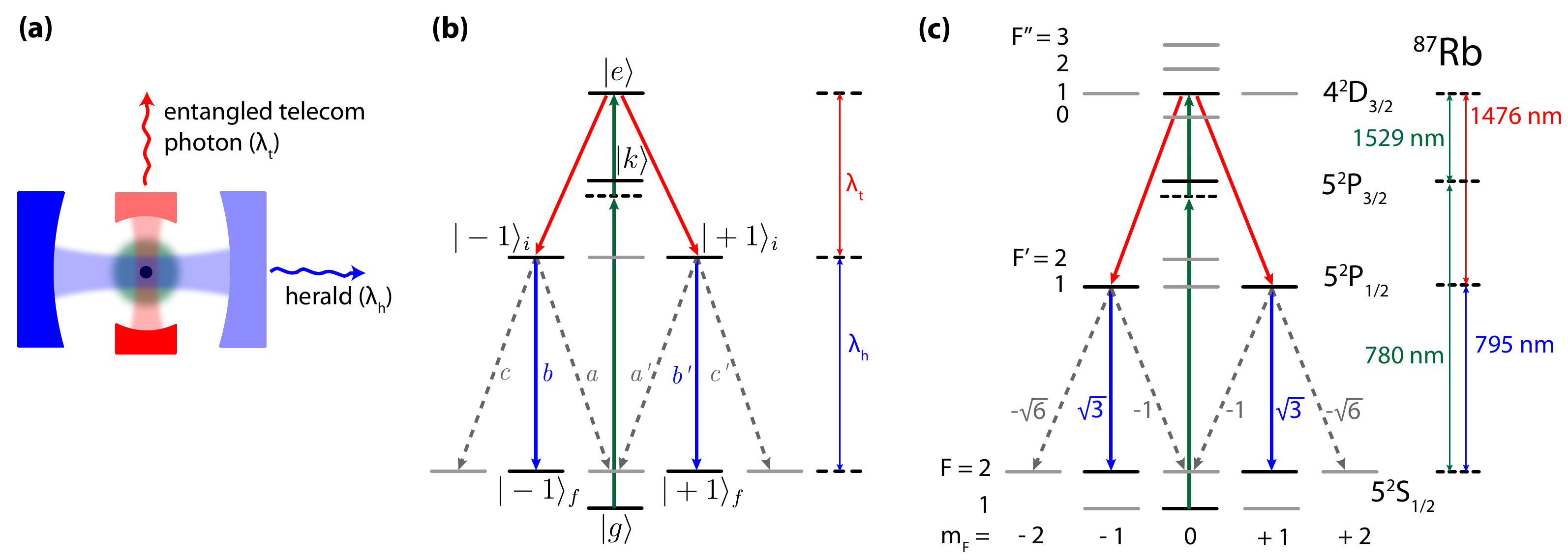}
\caption{\label{fig:SchemeEntanglementTelecom}
\textbf{a} Experimental setup for creating entanglement between a single atom and a telecom-wavelength photon. A single atom is placed at the intersection of the modes of a heralding cavity at $\lambda_\mathrm{h}$ and an entangling cavity at $\lambda_\mathrm{t}$. The cavities are single-sided, such that intracavity photons leave through the respective output-coupling mirror (lighter color) with high probability. The atom is driven on a two-photon transition via control beams impinging perpendicular to the image plane. \textbf{b} Atomic energy level scheme. The required states are labeled and drawn in black. Unwanted decay paths at $\lambda_\mathrm{h}$ are indicated by the gray dashed arrows. The relative transition amplitudes are indicated by the letters $a^{(\prime)}$ -- $c^{(\prime)}$. \textbf{c} A particular implementation with $^{87}$Rb. Zeeman states of hyperfine levels not relevant for the scheme have been omitted for clarity. The level spacings are not to scale.}
\end{figure*}

\section{Heralded atom-photon entanglement at telecom wavelength}
\label{sec:EntanglementScheme}
Direct coupling of an atom to single photons requires an atomic transition at the desired wavelength. Unfortunately, the atomic species that can be easily laser-cooled show no suitable transitions from the ground states at telecom wavelengths. However, there are suitable transitions from excited states, which have already been used for wavelength conversion \cite{Radnaev2010} and entanglement generation using cascade transitions \cite{Chaneliere2006} in atomic ensembles. Here, we present a related cascaded scheme that uses these transitions to generate entanglement between an atom and a telecom-wavelength photon without wavelength conversion.
We propose to use a single atom trapped at the intersection of two high-finesse cavities (see Fig.\,\ref{fig:SchemeEntanglementTelecom}a) such that it is coupled to two different, spatially and spectrally well-defined modes. This enables efficient emission of single photons at telecom wavelength entangled with the atom via one of the cavities and herald photons at near-infrared wavelength via the other. The herald photons signal the successful creation of entanglement.

\subsection{Cascaded entanglement scheme}
We assume an atom with a level scheme compatible with the one depicted in Fig.\,\ref{fig:SchemeEntanglementTelecom}b, which resembles the levels of bosonic alkali atoms. Essential for the protocol are two degenerate, long-lived states $\ket{\pm 1}_f$, which are Zeeman sublevels with $m_F=\pm 1$ and will be used to store the atomic qubit. These states are each coupled via a $\pi$-transition at wavelength $\lambda_\mathrm{h}$ to one of the two short-lived, intermediate states $\ket{\pm 1}_i$, which also feature $m_F=\pm 1$. The excited state $\ket{e}$, with $m_F=0$, can decay to $\ket{\pm 1}_i$ at wavelength $\lambda_\mathrm{t}$. The atom is initialized in a third ground state $\ket{g}$ with $m_F=0$, and can be coupled to $\ket{e}$ via a two-photon transition far detuned from an intermediate state $\ket{k}$, which may be identical to $\ket{i}$. The atom is placed at the intersection of two perpendicular cavity modes. The cavities are named heralding and entangling cavity and are resonant at $\lambda_\mathrm{h}$ and $\lambda_\mathrm{t}$, respectively. To describe the system, we choose the quantization axis to coincide with the axis of the entangling cavity. This way, the entangling cavity supports $\sigma^-$- and $\sigma^+$-polarization and the heralding cavity supports linear polarization modes, one of which needs to be aligned with $\pi$-polarization.

To create atom-photon entanglement, the atom is first initialized in state $\ket{g}$ by optical pumping and then coupled to the state $\ket{e}$ by a two-photon control pulse. Due to the presence of the resonant entangling cavity, this state couples cavity-enhanced to the states $\ket{-1}_i$ and $\ket{+1}_i$ and emits a photon in a superposition of the polarization states $\ket{\sigma^{+}}_t$ and $\ket{\sigma^{-}}_t$ into the entangling cavity. This controlled photon production is similar to entanglement generation schemes at near-infrared wavelengths \cite{Wilk2007a}. By varying the shape of the control pulse, the envelope of the generated photon can be controlled. If the process succeeds, the atomic state and the polarization of the photon leaving the entangling cavity are entangled:
\begin{equation}
\label{eq:intermediateState}
\ket{\Psi_1} = \frac{1}{\sqrt{2}}\left(\ket{-1}_i \ket{\sigma^{+}}_t + e^{\mathrm{i} \theta} \ket{+1}_i \ket{\sigma^{-}}_t\right).
\end{equation}
The relative phase $\theta$ can take the values 0 and $\pi$ and depends on the relative sign of the transition dipole matrix elements and thus on the particular transition chosen.

$\ket{\Psi_1}$ is already an entangled state between the atom and a photon at the wavelength $\lambda_\mathrm{t}$, but it is short lived, because the states $\ket{\pm1}_i$ quickly decay. If the decay leads to emission of a $\pi$-polarized photon, the entangled atom-photon state is transferred to the long-lived, final state
\begin{equation}
\label{eq:finalState}
\ket{\Psi_2} = \frac{1}{\sqrt{2}}\left(\ket{-1}_f \ket{\sigma^{+}}_t + e^{\mathrm{i} \tilde{\theta}} \ket{+1}_f \ket{\sigma^{-}}_t\right).
\end{equation}
Again, the relative phase $\tilde{\theta}$ can take values of 0 and $\pi$. Detection of a $\pi$-polarized photon at $\lambda_\mathrm{h}$ thus heralds the successful creation of the desired entangled state. To achieve the high success probabilities required for quantum repeaters, this decay path should dominate and the herald photons need to be efficiently collected. Both requirements are accomplished by the heralding cavity.

The proposed scheme is suitable for all bosonic isotopes of alkali atoms. Implementations with rubidium, cesium and francium are particularly interesting, because these elements have suitable transitions at wavelengths in telecom bands.

\subsection{Implementation with $^{87}$Rb}
\label{sec:Implementation}
To evaluate the performance of the proposed scheme with current technology, we investigate a particular implementation with $^{87}\mathrm{Rb}$ (Fig.\,\ref{fig:SchemeEntanglementTelecom}c) more closely. We choose the hyperfine states $\ket{F=1;m_F=0}$, $\ket{F=2; m_F=-1}$ and \linebreak[4] $\ket{F=2; m_F=+1}$ of the $5^2S_{1/2}$ state as $\ket{g}$, $\ket{-1}_f$, and $\ket{+1}_f$, respectively. As intermediate states $\ket{\pm1}_i$ the \linebreak[4] $\ket{F'=1;m_F=\pm1}$ substates of the $5^2P_{1/2}$ manifold are used and the $4^2D_{3/2}\ket{F''=1; m_F=0}$ state serves as the excited state $\ket{e}$. The $5^2P_{3/2}$ state can be employed as the intermediate state $\ket{k}$ for the two-photon control. This requires a heralding cavity resonant at $\lambda_\mathrm{h}=795$\,nm and will generate entangled photons at $\lambda_\mathrm{t}=1476$\,nm, which is in the S-band of optical fiber communication.

\begin{figure}
\includegraphics[width=\columnwidth]{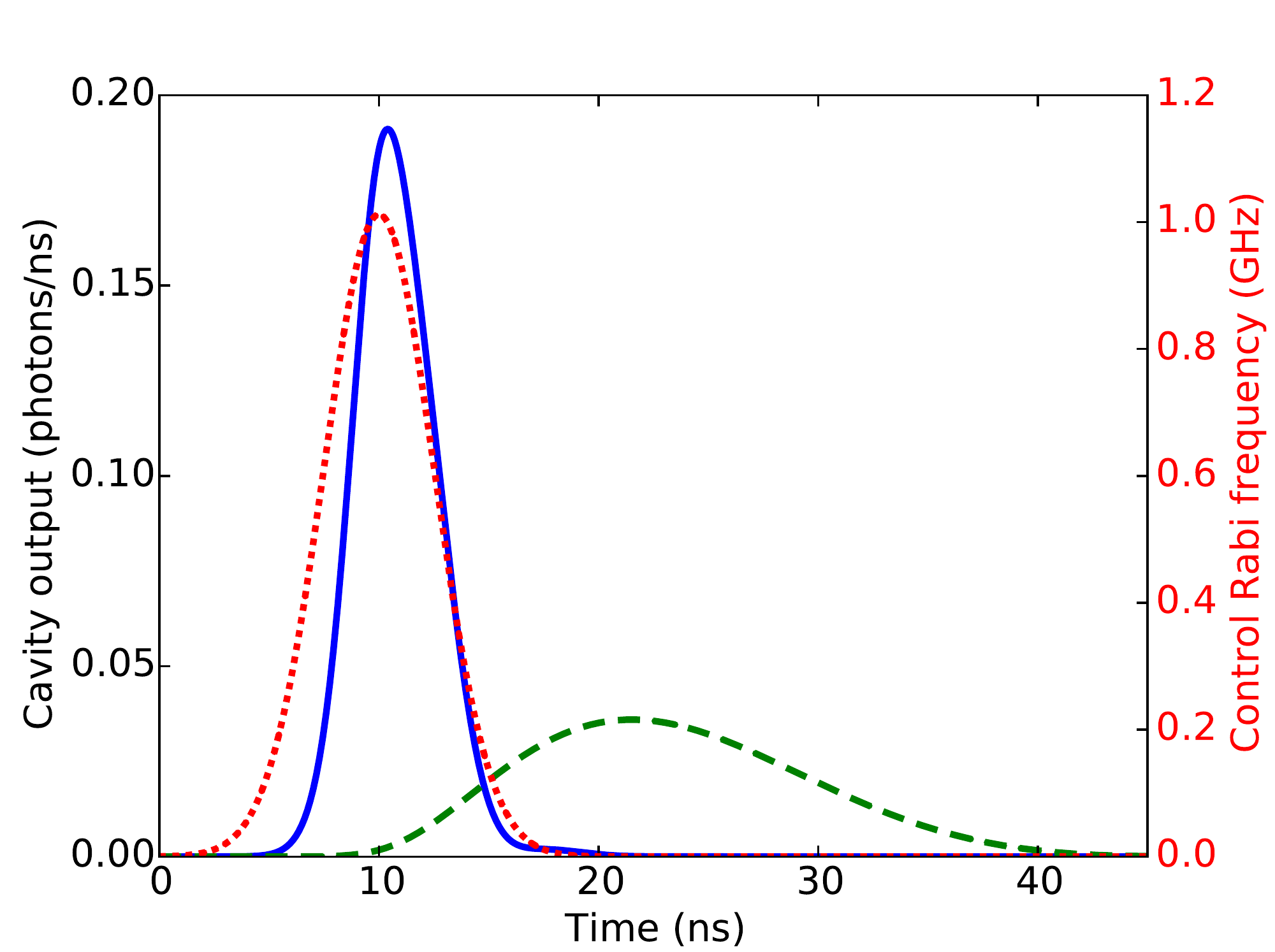}
\caption{Expectation value of the entangling-cavity output (blue solid line) and the heralding-cavity output (green dashed line) for the parameters given in the main text and a Gaussian control pulse (red dotted line) with a full width at half maximum (FWHM) of 5.9\,ns. Because the control pulse is short, the bandwidth of the photons is limited by the coupling strengths and the bandwidths of the cavities. Correlations of spectral and temporal properties between herald and telecom photons are therefore suppressed. At times beyond the 45\,ns plotted here, there is a small revival of the heralding cavity output, which, however, contains less than 3\,\% of the total output.}
\label{fig:SimulationMasterEq}
\end{figure}

\label{sec:NumericalValues}
The proposed scheme requires two cavities with small mode volumes and intersecting modes. We propose to use a Fabry-Perot cavity based on CO$_2$ laser-machined optical fibers \cite{Hunger2010} as the entangling cavity, because this type of cavity combines low mode volumes, high finesse and excellent optical access due to small external dimensions. Using telecom fibers as substrates also provides direct fiber integration, which facilitates long-distance communication. Fiber cavities can support degenerate polarization eigenmodes \cite{Takahashi2014}. This is assumed for the entangling cavity, such that the atomic state gets entangled with the polarization of the emitted telecom photon, but not with its frequency.
Starting from realistic cavity parameters (see Appendix \ref{app:Parameters} for details), we calculate an atom-light coupling rate of $g_\mathrm{t}=2\pi\times 70\,\mathrm{MHz}$, a cavity field decay rate through the outcoupling mirror of $\kappa_\mathrm{t}^\mathrm{oc}=2\pi\times 95\,\mathrm{MHz}$, and an additional cavity decay caused by other losses of $\kappa_\mathrm{t}^\mathrm{l}=2\pi\times 8\,\mathrm{MHz}$ for the entangling cavity. Photons leaving this cavity through the output coupler couple to the mode of a standard single-mode telecom fiber with an efficiency of 0.96. 

For the heralding cavity, fiber integration is not necessary, and we therefore propose to use CO$_2$ laser-machined glass plates as mirror substrates \cite{Petrak2011}. Degeneracy of its polarization eigenmodes is undesirable, because the mode with polarization orthogonal to $\pi$ can enhance wrong decay paths (shown as dashed arrows in Fig.\,\ref{fig:SchemeEntanglementTelecom}). This is detectable, because it does not result in a $\pi$-polarized herald photon. Consequently, the failed entangling attempt is discarded, thereby reducing the efficiency. To avoid this problem, we suggest to employ a heralding cavity with a large frequency splitting of the polarization eigenmodes \cite{Uphoff2015}, such that the mode orthogonal to $\pi$-polarization is far detuned from the atomic transition. We assume the heralding cavity parameters to be $\{g_\mathrm{h},\kappa_\mathrm{h,}^\mathrm{oc},\kappa_\mathrm{h}^\mathrm{l}\}=2\pi\times\{16.3, 11.9, 1.5\}\,\mathrm{MHz}$. The total cavity decay rates are given by $\kappa_\mathrm{t}=\kappa_\mathrm{t}^\mathrm{oc}+\kappa_\mathrm{t}^\mathrm{l}$ and $\kappa_\mathrm{h}=\kappa_\mathrm{h}^\mathrm{oc}+\kappa_\mathrm{h}^\mathrm{l}$. The cooperativity of the atom-cavity system is $C_\mathrm{t (h)} = g_\mathrm{t(h)}^2\slash\left(\kappa_\mathrm{t(h)} \Gamma_\mathrm{t(h)}\right) = 25~(3.4)$ for the entangling (heralding) atom-cavity system. Here, $\Gamma_\mathrm{t} = 2\pi \times1.92\,\mathrm{MHz}$ $(\Gamma_\mathrm{h}=2\pi \times 5.75\,\mathrm{MHz})$ is the decay rate of the $4^2D_{3/2}$ $(5^2P_{1/2})$ state of $^{87}$Rb. 

\begin{figure}
\includegraphics[width=\columnwidth]{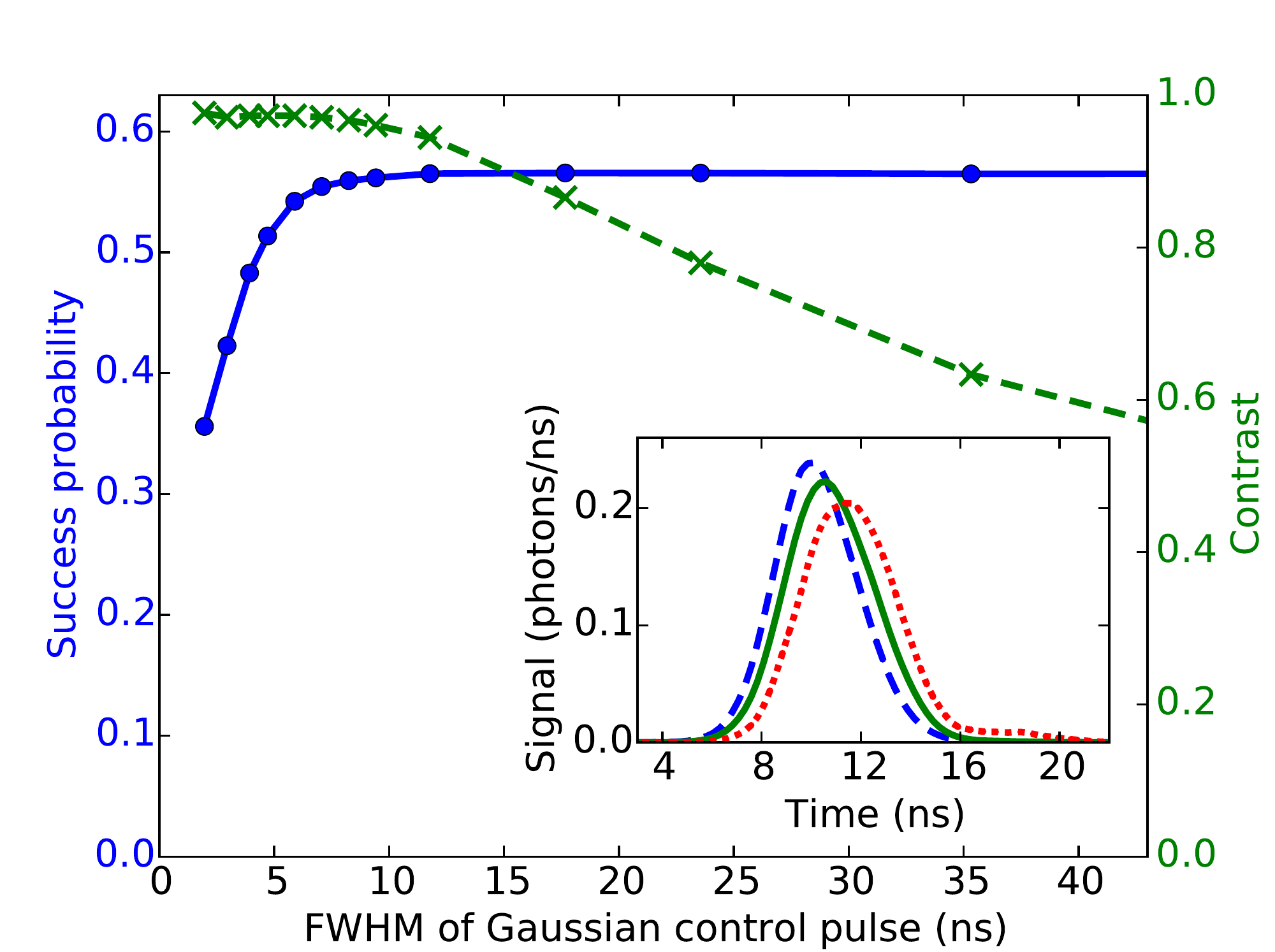}
\caption{Overall success probability (blue dots and solid line, left axis) and estimated contrast (green crosses and dashed line, right axis, see Sec. \ref{sec:Indistinguishability}) for Gaussian control pulses of different widths. The amplitude of the control pulses has been set to the minimum value that leaves less than 1\,\% population in the initial state. Dots and crosses are the results of the numerical simulations, which have been connected by lines to guide the eye. The inset shows the expectation value of the entangling-cavity output as a function of time, conditioned on a photon leaving the heralding cavity after 15\,ns (i.e., early, blue dashed line), 22\,ns (green solid line) and 35\,ns (i.e., late, red dotted line) for a control pulse with a FWHM of 5.9\,ns (as in Fig.\,\ref{fig:SimulationMasterEq}). There is a weak correlation between the arrival time of the herald photon and the telecom-photon envelopes, leading to a slightly reduced contrast.}
\label{fig:EfficiencyContrast}
\end{figure}

We perform numerical simulations on this system using two methods. The first one is integration of the corresponding Lindblad master equation, from which we extract the independent expectation values for the heralding- and entangling-cavity output. In the situation depicted in Fig.\,\ref{fig:SimulationMasterEq}, the control pulse is long enough to result in a near-Gaussian shape of the telecom photons, but short enough that the bandwidth of the telecom and herald photons is limited by the respective coupling and cavity decay rates. Because the timescale for the decay of the heralding cavity is different from that of the entangling cavity, the arrival time of the herald photons is determined by the properties of the heralding cavity and only weakly correlated with the arrival time of the telecom photon. This enables the emission of near-indistinguishable telecom photons (inset of Fig.\,\ref{fig:EfficiencyContrast}). 

The second method is a Monte Carlo wave-function approach \cite{Dalibard1992,Dum1992,Tian1992}, which yields information about correlations between the cavity outputs. From the latter, we calculate the overall success probability $p_\mathrm{ht}$, i.e., the probability to obtain a telecom photon in the optical fiber and a herald photon leaving the heralding cavity through the output coupler. For the parameters chosen here, we find $p_\mathrm{ht}=0.57$, basically independent of the length of the control pulse (see Fig.\,\ref{fig:EfficiencyContrast}). The exception are very short control pulses, for which we calculate lower success probabilities, because the control pulse gets spectrally broad enough to excite the $4^2D_{3/2}\ket{F''=3; m_F=0}$ state. The main loss channels are spontaneous decay of the $5^2P_{1/2}$ state (probability 0.24) and parasitic losses in the entangling and heralding cavity (probability 0.08 and 0.07, respectively). By increasing the coupling between atom and heralding cavity and decreasing the parasitic losses in both cavities, these loss channels could be minimized. Thus, with improvements in technology, the scheme could be performed almost deterministically.

\subsection{Fidelity}
At the start of the protocol the atom needs to be initialized in $\ket{g}$. This can be achieved by optical pumping, which is susceptible to experimental imperfections. Because there is no fundamental limit and various strategies are possible, we assume that the atom is initialized in $\ket{g}$ with unit efficiency and the control pulse addresses the state $\ket{e}$ without exciting any other state. Under these conditions, the fidelity of the intermediate atom-photon state with the ideal state $\ket{\Psi_1}$ is determined solely by the geometry of the system and we assume any photon leaving the entangling cavity to be in the correct entangled state. This is a very good approximation if the $5^2P_{1/2}$ state of $^{87}$Rb is used as the intermediate state, because undesired processes are far off-resonant. The transition from the intermediate to the final state is unambiguously heralded by a photon with the correct polarization at wavelength $\lambda_\mathrm{h}$. Therefore, conditioned on the detection of a herald photon, high-fidelity entanglement generation should be possible. In the following, we study two potentially detrimental effects and demonstrate that their influence on the fidelity of the entangled state is marginal for the particular implementation and parameters chosen here.

\subsubsection{Second polarization mode of the heralding cavity}
\label{sec:secondMode}
Degenerate polarization eigenmodes of the heralding cavity reduce the efficiency of the protocol. Besides this obvious effect, the unwanted resonant mode of the herald cavity also has a subtle but detrimental impact on the fidelity, because it leads to a state-dependent probability for herald generation (cf. Appendix \ref{app:secondMode}). 

To see this, we rewrite $\ket{\Psi_1}$ in a linear polarization basis for the telecom photon:
\begin{eqnarray}
\label{eq:intermediateStateBasisTrafo}
\ket{\Psi_1} = \frac{1}{2} \left[ \ket{H}_t  \left(\ket{-1}_i + e^{\mathrm{i} \theta} \ket{+1}_i \right) \right. \nonumber \\
+ \left. \ket{V}_t  \left(\ket{-1}_i - e^{\mathrm{i} \theta} \ket{+1}_i\right) \right].
\end{eqnarray}
The two atomic states correlated with a horizontally or vertically polarized photon, respectively, are coupled differently to the $m_F=0$ sublevel of the final state $\ket{f}$ by the second mode of the heralding cavity. The transition amplitudes from one of the two superpositions $(\ket{-1}_i\pm\ket{+1}_i)\slash\sqrt{2} $ will interfere destructively, while the other interferes constructively. If the second mode of the heralding cavity with a polarization orthogonal to $\pi$-polarization is resonant as is the case for a cavity with degenerate polarization eigenmodes, these superpositions have different probabilities to emit a $\pi$-polarized photon, because one of these cannot decay to the state with $m_F=0$ via the herald cavity, while the other one can. Therefore, the final state, postselected on a correctly polarized herald photon, is not exactly $\ket{\Psi_2}$, but has unequal amplitudes and is consequently not a maximally entangled state. To prevent that and thereby maximize the fidelity, the atom should not couple to the second mode.

In the implementation with $^{87}$Rb (Sec.\,\ref{sec:Implementation}), the effect is small, because the unwanted mode of the heralding cavity couples only weakly to the state with $m_F=0$ compared to the coupling to the states with $m_F\neq0$. This is due to the specific branching ratios of the states involved. Even if the orthogonal polarization mode is degenerate with the $\pi$-polarized heralding mode, the reduction in fidelity is only 0.15\,\% (see Appendix \ref{app:secondMode}).  The effect is minimized if there is a large frequency splitting between the polarization eigenmodes, as also required to maximize the efficiency.

\subsubsection{Free-space decay}
If the atom decays via free-space emission, a state different from the desired one might be created. If no herald photon is emitted, the resulting state can be discarded. This results in a reduced efficiency but leaves the process fidelity unaffected. But, if the intermediate states $\ket{\pm1}_i$ can decay back to the initial state $\ket{g}$ or any other state that can be excited by the control, there is a chance for multi-photon events. Reexcitation of $\ket{e}$ can result in a second telecom photon, which leaves the system in an undesired entangled state, but can still result in the generation of a herald. There are two ways to minimize the generation of multiple telecom photons. The first one is to choose the intermediate states in such a way that spontaneous decay preferably puts the atom in states that are not excited by the control laser. The second one is to execute the scheme quickly on the time scale of atomic decay. In the limit of fast excitation \cite{Maunz2007}, for example with a picosecond laser, the detrimental effect would be eliminated. However, this would remove the ability to influence the shape of the photonic wave packet with the control lasers and a broadband pulse might excite other states, which do not couple to the entangling cavity and thus reduce the efficiency.

We employ the Monte Carlo wave-function approach to calculate an upper bound for multi-photon events for the parameters of Sec.\,\ref{sec:NumericalValues} and a short Gaussian control pulse of 5.9\,ns FWHM. We assume the worst case, namely that all decays to states outside the simulated system will result in a decay to the initial state, from which the atom can be efficiently excited again by the control laser. We compare the number of calculated quantum trajectories which result in multiple photons being generated to the number of desired trajectories and find that less than 0.4\,\% of the herald photons are accompanied by more than one telecom photon. More realistically, the atom will also decay to other states, which might not be excited as efficiently as the initial state, such that the number of multi-photon events should be lower than this upper bound.

\section{Remote entanglement of two systems}
\label{sec:RemoteEntanglement}
The creation of remote entanglement between two of the quantum-repeater nodes described in the previous section can be achieved by an optical Bell-state measurement (BSM) \cite{Zukowski1993,Weinfurter1994} on the polarization of the two telecom-wavelength photons that are entangled with the atom at their respective repeater node. The BSM is based on two-photon interference and therefore requires the telecom photons to be indistinguishable in their temporal, spectral, and spatial properties, despite their remote origins. Single atoms in optical cavities can be controlled to exactly defined conditions, enabling indistinguishable photon-generation processes at remote locations \cite{Nolleke2013}. However, possible temporal correlations of the telecom photon with the herald photon have to be considered. Emission via a two-photon cascade in free space shows a clear order of the photons. Emission of a first photon from the upper part of the cascade is followed by an exponential decay of the photon belonging to the lower part. Conditioned on the arrival time of the second photon, the first photon will have an exponentially rising envelope, with a sharp drop to zero at the detection time of the second photon \cite{Gulati2014}. Telecom photons corresponding to herald photons detected at different times will thus have different arrival-time distributions, which renders the photons distinguishable. Postselecting on events where the herald photons were detected with the same delay relative to the control pulse would reestablish indistinguishability, but also severely limit the efficiency.

The cavity for the herald photon offers a way out: If the lifetime of the heralding cavity is very long compared to the wave-packet envelope of the generated telecom photon, the former will determine the wave packet and therefore the detection-time distribution of the herald photon. Consequently, the correlations between the detection time of the herald photon and the wave-packet shape of the telecom photon will be erased by the heralding cavity, thereby restoring indistinguishability between telecom photons of different origin. The ideal implementation requires an entangling cavity with large coupling between atom and telecom photon and large cavity decay rate. The heralding cavity should have a smaller decay rate, and a smaller coupling between atom and herald photon can be tolerated.
	
\subsection{Numerical calculation of the photon indistinguishability}
\label{sec:Indistinguishability}
In order to quantify the indistinguishability of the telecom photons, we use the Monte Carlo wave-function approach to generate arrival time pairs of telecom and herald photons. These pairs are used as input to a kernel density estimator in order to estimate the probability distribution of telecom photons conditioned on a photon leaving the heralding cavity at a specific time. As the kernel, we use a two-dimensional Gaussian function with the bandwidth along the two axes set to $6\kappa_\mathrm{t}$ and $6\kappa_\mathrm{h}$, respectively. This bandwidth should be larger than the bandwidths of the processes occurring in the cavities, so the resulting probability distribution is likely undersmoothed. The resulting telecom-photon probability, conditioned on the detection time of the herald photon, is shown in the inset of Fig.\,\ref{fig:EfficiencyContrast} for the same parameters used in the simulation depicted in Fig.\,\ref{fig:SimulationMasterEq}. Under the assumption that the processes at remote locations are identical in all other aspects, we calculate the interference contrasts expected in a Hong-Ou-Mandel experiment \cite{Hong1987,Nolleke2013} for the telecom photons conditioned on photons leaving the heralding cavity at different times. Weighting the resulting contrasts with the probability distribution for the herald photons yields the average contrast $C$, which is 0.97 in this case. This two-photon interference contrast can be converted into a remote-entanglement fidelity $F= \frac{1}{2}(1 + C)$ under the assumption that all other processes are perfect \cite{Rosenfeld2011a}. Thus, a fidelity of close to 0.99 should be achievable with our model parameters.

We repeat the calculation for Gaussian control pulses of different width (Fig.\,\ref{fig:EfficiencyContrast}). For long control pulses, the emission time of the herald photon is correlated with the emission time of the telecom photon, which reduces the indistinguishability and thus the interference contrast of the telecom photons. For very short control pulses, there is little correlation and the interference contrast is therefore high; the success probability, however, is not maximal. Both the success probability and the interference contrast are near their respective maximum for control pulses with a FWHM between 5\,ns and 10\,ns. This is therefore an ideal point of operation, with minimal sacrifices in the tradeoff between efficiency and fidelity. Because of the use of undersmoothed probability distributions, the values for the contrast are limited by the variance of the Monte Carlo method and therefore present a lower limit to the theoretically achievable contrast.

The width of the envelope of telecom and herald photons is well above the timing resolution of commercially available single-photon counters. It is therefore possible to postselect events using the arrival times of telecom and herald photons. This enhances the interference contrast and thereby the entanglement fidelity, however at the cost of a reduced total success probability. In the limit of identical detection time of the herald photons and in the absence of any other effect that makes the telecom photons distinguishable, an interference contrast of unity is reached.

\section{Entanglement swapping}
\label{sec:EntanglementSwapping}
To connect separate repeater links, entanglement swapping has to be performed. This requires a BSM on two quantum memories, each of which is entangled with a remote node. The small size of the fiber-based entangling cavities enables the placement of two of them in one and the same heralding cavity (Fig.\,\ref{fig:Repeater}). The heralding cavity can then be used for the collection of herald photons from both entangling cavities. To this end, the creation of atom-photon entanglement has to be alternated between the two atoms, which is possible by addressing the atoms with individual control beams and selective detuning, e.g., via a local light shift. The remote entanglement procedure described in the previous section can thus be repeated individually for each atom until it has succeeded for both.

A quantum repeater requires quantum memories with long coherence times. The coherence between the Zeeman states $\ket{-1}_f$ and $\ket{+1}_f$ is limited by fluctuations of the effective magnetic field \cite{Rosenfeld2011}. States with spin-orbit angular momentum $J=1/2$, different hyperfine quantum number and the same magnitude but opposite sign of the magnetic quantum number feature a reduced differential Zeeman shift. Consequently, coherent superpositions of these states have a strongly reduced sensitivity to magnetic field fluctuations \cite{Harber2002}. We therefore propose to use a microwave pulse to state-selectively transfer one of the Zeeman states (e.g., $\ket{-1}_f$) to the other hyperfine ground state with the same $m_F$ with high fidelity. The corresponding final qubit states of $^{87}\mathrm{Rb}$ are $\ket{F=1;m_F=-1}$ and $\ket{F=2;m_F=+1}$. At a moderate magnetic field of about 3.23\,G, the two atomic states experience the same first-order Zeeman shift and a coherence time of several seconds has been observed \cite{Treutlein2004,Deutsch2010}.

As the heralding cavity can couple to both atoms, it is a natural choice for the implementation of an interaction mechanism for entanglement swapping \cite{Sorensen2003,Duan2004,Borregaard2015a}. We propose to use a quantum gate between the two atoms based on the reflection of a resonant single photon from the cavity \cite{Duan2004,Reiserer2014}. As only one of the two hyperfine ground states couples to the heralding cavity, this results in a state-dependent phase shift of $\pi$ on the two-atom state, equivalent to a controlled-Z quantum gate \cite{Duan2005}.

To perform entanglement swapping, we start with a Hadamard single-qubit rotation on one of the atoms, then apply the controlled-Z gate and subsequently a Hadamard gate on each of the two atoms. This maps the four atomic Bell states unambiguously onto four separable atomic states. These can be detected with unity efficiency and high fidelity by performing cavity-assisted hyperfine state detection \cite{Gehr2010,Bochmann2010} on each of the atoms. This entanglement swapping projects the remote nodes into one of the four Bell states. The measurement result identifies the created Bell state. Therefore, conditional single-qubit operations at the remote nodes might then be used to rotate the entangled state into a specific target state. Although realistic implementations of reflection-based gates have a failure probability, the success of the gate is heralded by the detection of the reflected photon. The method requires a single-sided cavity and equal reflectivities of the empty cavity and the coupled atom-cavity system. These requirements are compatible with the design criteria of the heralding cavity posed by the entanglement scheme presented in Sec.\,\ref{sec:EntanglementScheme}.

\section{Quantum-repeater performance}
\begin{figure}
\includegraphics[width=\columnwidth]{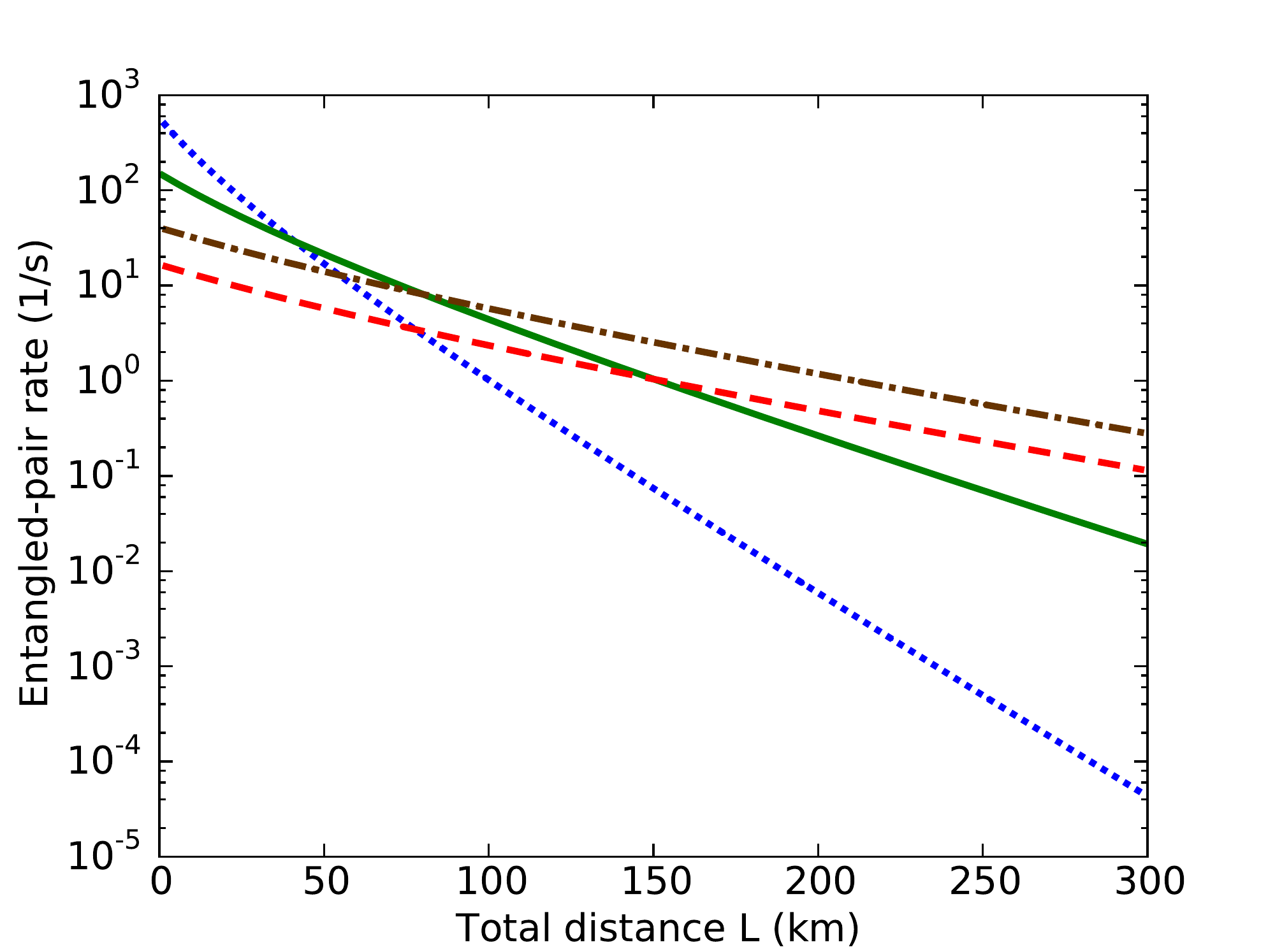}
\caption{Semilogarithmic plot of the expected average rate of heralded entangled pairs between two end points separated by the total distance $L$ with one elementary link (i.e., no repeater, blue dotted line), two links (green solid line), four links with restart from beginning (red dashed line) and four links keeping already entangled pairs (brown dash-dotted line). Because of the lower overhead, the scenario with no repeater node yields the highest rates at short distances, but scales worse with $L$. For $L \geq 100$\,km the quantum repeater protocols are clearly superior. See the main text for the parameters used in the calculations.}
\label{fig:RepeaterRates}
\end{figure}

\label{sec:Repeater}
The techniques described in the previous sections can be used to implement a quantum repeater protocol with entanglement generation, entanglement distribution and entanglement swapping. To assess the performance of such a repeater, we consider two atoms trapped in remote cavities separated by the total distance $L$, which are to be entangled. We calculate the rate at which heralded entangled pairs are produced at these end points with the help of repeater nodes separated by the distance $L_0$ and compare it to the rate achievable without repeater nodes. If all processes succeed with high fidelity, fidelity differences between the protocols are marginal and the entangled pair rates can be compared directly. In the case of experimental imperfections, the reduced fidelity has to be compensated by a higher entangled-pair rate, e.g., to generate a secret key of the same length (see Sec. \ref{sec:secretKeyRate}).
\subsection{Entangled-pair rate}
The probability to successfully generate remote entanglement in a single attempt between two adjacent nodes separated by distance $L_0$ is
\begin{equation}
p_\mathrm{e}(L_0) = \frac{1}{2}(p_\mathrm{ht} \eta_\mathrm{h} \eta_\mathrm{t} )^2 \exp(-L_0/L_\mathrm{a}),
\end{equation}
where $p_\mathrm{ht}$ is the success probability of the entanglement generation, $\eta_\mathrm{h}$ and $\eta_\mathrm{t}$ are the detector efficiencies at herald and telecom wavelength, respectively, and $L_\mathrm{a}$ is the attenuation length of the fiber. The factor $1/2$ results from the fact that with a photonic BSM, only two of the four Bell states can unambiguously be identified. The average number of attempts required for $N$ parallel processes with individual success probability $p$ to have each succeeded once is \cite{Bernardes2011}
\begin{equation}
Z_\mathrm{N}(p) = \sum_{j=1}^{N}{N\choose{j}}\frac{(-1)^{j+1}}{1-(1-p)^j}.
\end{equation}
It thus on average takes $Z_\mathrm{N}(p_\mathrm{e}(L_0))$ attempts to generate $N$ entangled pairs. The atomic BSM is not completely deterministic, but succeeds with probability $p_\mathrm{es} = R p_\mathrm{p} \eta_\mathrm{h}$, where $R$ is the reflectivity of the herald cavity and $p_\mathrm{p}$ is the efficiency of the single photon source that generates the photon to be reflected. Therefore, we consider two different strategies for entanglement swapping. The first one is to swap all entangled pairs at once and restart the whole protocol when one entanglement swapping attempt fails (with probability $1-p_\mathrm{es}$). In that case, the average number of attempts $\left<n(N)\right>$ until the protocol has succeeded over the distance $L=N L_0$ is
\begin{equation}
\left<n(N)\right> = \frac{Z_\mathrm{N}(p_\mathrm{e}(L_0))}{p_\mathrm{es}^{N-1}}.
\end{equation}

The second strategy is to swap entangled pairs as soon as possible. Upon failure, all entangled pairs that were not part of the failed entanglement-swapping attempt are kept and entanglement between the others is reestablished. Obviously, for $N>2$ this results in a higher rate of entangled pairs, but the memories also need to store the entangled states for a longer time, leading to increased decoherence. To calculate $\left<n\left(N\right)\right>$ for this strategy, we perform a Monte Carlo simulation with $10^6$ runs per calculated point.

The total time $\left<T\right>$ that is on average required to entangle the endpoints can be immediately calculated:
\begin{equation}
\left<T\right> = \left<n(N)\right> \left(\frac{L_0}{c_\mathrm{f}} + \tau\right).
\end{equation}
Here, $c_\mathrm{f}$ is the speed of light in the optical fiber and $\tau$ is the minimum latency between two attempts, which is expected to be dominated by the time required for cooling and optical pumping of the atom.

To be quantitative, we take $p_\mathrm{ht}$ as calculated in Sec.\,\ref{sec:NumericalValues} for a control pulse with 5.9\,ns FWHM, which combines high efficiency with high indistinguishability as estimated in Sec.\,\ref{sec:Indistinguishability}. Correcting this value for the slightly reduced coupling of atoms not in the center of the herald cavity yields $p_\mathrm{ht} = 0.53$ (see Appendix \ref{app:Parameters}), which is the value we use throughout this section. 
$L_\mathrm{a} = 22\,\mathrm{km}$ and $c_\mathrm{f} = 2\cdot10^5\,\mathrm{km\slash s}$ are typical parameters for commercial telecom optical fibers. We assume detectors with an efficiency of $\eta_\mathrm{t} = \eta_\mathrm{h} = 0.8$ which is within range of current technology \cite{Marsili2013}. The reflectivity of the cavity is $R=0.61$ for the considered heralding cavity (Appendix \ref{app:Parameters}). Single atoms trapped in optical cavities have been shown to be highly efficient single-photon sources \cite{Mucke2013}, and with state-of-the-art cavity parameters they could certainly achieve an efficiency of $p_\mathrm{p}=0.8$, which is the value we assume for the single photon source used in the atomic BSM. We further consider a conservative cycle time of $\tau = 100$\,\textmu s. This would result in a repetition rate of 10\,kHz over very short distances. For larger separations, the finite $c_\mathrm{f}$ and the resulting communication time have to be accounted for, which reduces the repetition rate. For example, at $L_0=80$\,km the repetition rate drops to 2\,kHz, dominated by the communication time.

\begin{figure}
\includegraphics[width=\columnwidth]{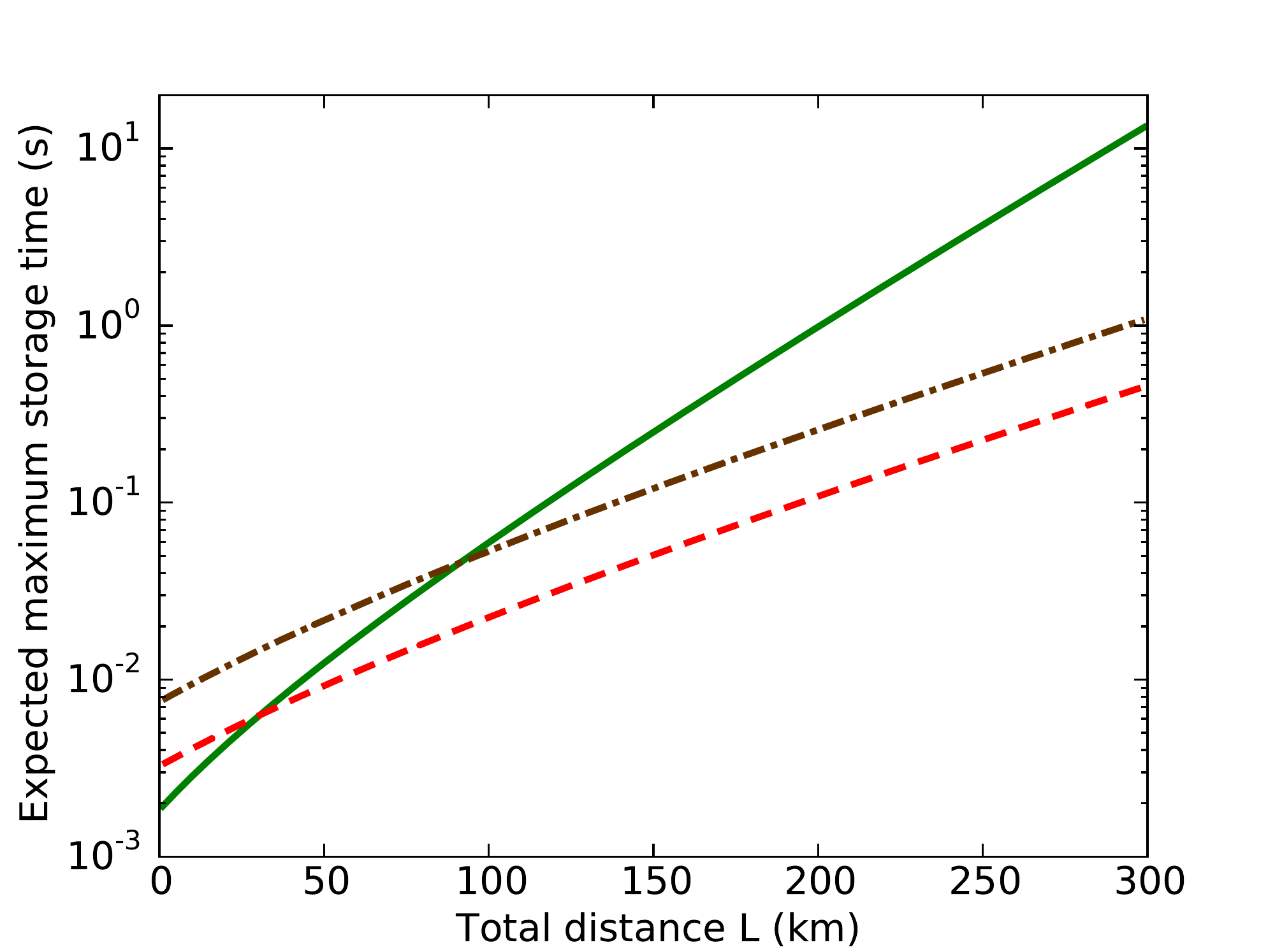}
\caption{Average storage time required for the nodes which have to retain entanglement the longest to entangle two end points separated by the total distance $L$. The coherence time of the nodes needs to greatly exceed this time to prevent degradation of the entanglement. The line styles are the same as in Fig.\,\ref{fig:RepeaterRates}. Repeaters with more nodes show a better scaling of the required storage time with distance, but have larger overhead. With four links, restarting from the beginning significantly relaxes the required storage time.}
\label{fig:RepeaterStorageTime}
\end{figure}

Using the above-mentioned parameters, we calculate the rate $1/\left<T\right>$ for the generation of remote heralded entangled pairs, without repeater ($N=1$), for the most basic repeater ($N=2$), and for the extension to the second swap level ($N=4$). This entangled-pair rate is plotted as a function of the total length of the repeater link in Fig.\,\ref{fig:RepeaterRates}. For $N=4$ we also compare the two different entanglement swapping strategies explained above. At short distances, direct entanglement is superior because of the reduced overhead compared to the repeater protocol. The break-even with one repeater node is reached after 41\,km, and after 100\,km, the repeater-assisted entanglement generation is approximately 4 times faster. A system with three repeater nodes with restart from the beginning outperforms the single-repeater-node system beyond 150\,km and is 14 times faster than direct entanglement at this distance. By keeping entangled pairs, another factor of approximately 2.4 can be gained and the break-even with the single-repeater-node system is already reached at 82\,km separation.

\subsection{Required storage time}
To prevent significant degradation of the entanglement, the coherence time of the employed quantum memories has to be much longer than the time the atomic state has to be stored. We calculate the average time interval between the first successful creation of entanglement and success of the protocol, which is the time the entanglement has to be stored (Fig.\,\ref{fig:RepeaterStorageTime} and Appendix \ref{app:RequiredStorageTime}). More repeater nodes result in better scaling with distance but have a larger overhead. Keeping unaffected entangled pairs after a failed entanglement-swapping attempt requires 2.4 times longer storage times.
At distance $L=100$\,km ($L=200$\,km) the single-repeater-node protocol requires a storage time of 59\,ms (980\,ms). Using three repeater nodes this value drops to 53\,ms (260\,ms). By restarting the protocol from the beginning, once an entanglement swapping attempt fails, the required storage time for three repeater nodes can be further reduced to 22\,ms (110\,ms). Aborting long-running attempts can thus be used to extend the maximum distance of a repeater system under the constraint of a limited coherence time, however at the cost of a constant factor in the entangled-pair rate. If memory time is a strong limitation, it can make sense to add more repeater nodes than Fig. \ref{fig:RepeaterRates} suggests to be optimal. Coherence times of several seconds have been demonstrated for the suggested qubit implementation in $^{87}$Rb and could be further improved by reducing magnetic field fluctuations \cite{Treutlein2004,Deutsch2010}. The degradation of the fidelity of the entangled pairs will, however, be a relevant limitation for the fidelity of the final state and therefore poses a challenge for any experimental implementations.
\subsection{Secret-key rate}
\label{sec:secretKeyRate}
An especially important application of a quantum repeater is quantum key distribution (QKD). Device-independent QKD is enabled by heralded entanglement between remote nodes. The raw key extracted from an imperfect entangled state can be converted into an unconditionally secret key by classical post-processing. In the limit of infinitely long keys,  the secret-key rate can be obtained by multiplying the entangled pair rate with the secret fraction of the final state \cite{Scarani2009}. The latter can be calculated from the indistinguishability of the telecom photons characterized by their interference contrast and the fidelity of an entangled state produced by the atomic BSM with perfect input states (see Appendix \ref{app:SecretFraction}). We assume that the coherence time of the memories is so long that we can neglect degradation due to decoherence. For an interference contrast of $C=0.97$, the BSM fidelity needs to be 95\,\% (89\,\%) to yield a secret fraction of 0.5 (0.25). As long as the fidelity is above 83\,\%, a secret key can be extracted. 
The secret fraction decreases if additional elementary links are inserted. With four elementary links and $C=0.97$ interference contrast, the BSM fidelity needs to be 99\,\% (97\,\%) to retain a secret fraction of 0.5 (0.25). In this case, the fidelity needs to be above 95\,\% for a non-zero secret fraction.

\subsection{Entanglement purification}
\label{sec:Purification}
Imperfections in the sub-steps required for the operation of a quantum repeater lead to an error in the final state that scales exponentially with the number of nodes. This can be overcome via entanglement purification \cite{Briegel1998}, which uses multiple states that are not maximally entangled to generate a single entangled state with higher fidelity. The purification protocol described by Deutsch et al. \cite{Deutsch1996} could be implemented in the repeater architecture proposed here by employing the single-qubit rotations, the atom-atom gate mechanism, and atomic state detection procedures proposed for entanglement swapping (see Sec. \ref{sec:EntanglementSwapping}). The quantum gates between two atoms in the same herald cavity would be complemented by a remote gate between two atoms in spatially separated herald cavities, also mediated by the reflection of a single photon \cite{Duan2005}, thereby enabling multiple rounds of entanglement purification using several instances of the proposed repeater node.

Entanglement purification can only increase the fidelity of the final state if the errors introduced by the required operations are significantly smaller than the errors present in the initial states. If entanglement purification and entanglement swapping build on the same gate mechanism, they will suffer from similar imperfections. For the very few entanglement swapping gates necessary for two and four elementary links, entanglement purification can therefore only lead to an increase in fidelity if considerable errors are introduced by other parts of the repeater protocol. An important source of error is loss of interference contrast $C$ in the entanglement distribution process, caused by either the generation of partially distinguishable photons or errors introduced during fiber transmission. Taking an interference contrast of $C=0.97$ (see Sec. \ref{sec:Indistinguishability}) as an example and applying the error models described in Appendix \ref{app:SecretFraction}, we find that the entanglement purification protocol described by Deutsch et al. \cite{Deutsch1996} increases the fidelity of the final state only if the atom-atom gate error is less than 4\,\% for two elementary links ($N=2$) and less than 7\,\% for four elementary links ($N=4$). 

In the case of perfect gates, any target fidelity below unity can in principle be reached. This comes, however, at the cost of a significant additional overhead, because more quantum memories are required, and the final entangled pairs are produced at a lower rate. For applications in quantum key distribution, the secret-key rate is a convenient measure that allows comparing repeater strategies with and without entanglement purification. Employing the methods described in Appendix \ref{app:SecretFraction}, we calculate that even with perfect gate operations, i.e., efficiency and fidelity of unity, entanglement purification is beneficial only if the interference contrast is very low, namely $C\leq 0.55$ in case of $N=2$ and $C\leq 0.83$ in case of $N=4$. This has to be compared to our expected interference contrast of $C=0.97$ (see Sec. \ref{sec:Indistinguishability}) and the fact that high-fidelity transport of polarization qubits in optical fibers over the distances considered here is possible \cite{Hubel2007}.

We therefore conclude that for a realistic repeater implementation over a few hundred kilometers, entanglement purification is unlikely to be beneficial and the best strategy is to focus on high-fidelity implementations without entanglement purification \cite{Sangouard2011,Bratzik2013}. 
 
\section{Conclusion}
Past experiments with single atoms in optical cavities have underlined their excellent prospects for applications in quantum communication \cite{Ritter2012,Nolleke2013,Reiserer2014}. The distances that can be bridged by direct photon transmission at near-infrared or visible wavelengths is, however, limited. We have shown how this limit could be overcome by employing, first, operation at telecom wavelength and, second, a realistic, efficient, and highly-integrated quantum-repeater concept.

Operation at telecom wavelength is essential for long-distance fiber-optic communication without the need for wavelength conversion. Our simulations show that creation of entanglement between single atoms and telecom-wavelength photons with an efficiency of 0.57 is possible with $^{87}$Rb atoms and current cavity technology. The designated S-band transition enables integration into existing fiber-optic networks. Remote entanglement can be established by a photonic BSM with a state fidelity of 99\,\%, because the generated photons are highly indistinguishable. Very efficient entanglement swapping can be achieved by a two-atom quantum gate employing the heralding cavity. Thus entanglement generation, quantum memories and entanglement swapping can be realized with single atoms in optical cavities, and all additional components necessary to complete a simple quantum repeater are commercially available. The highly efficient, yet heralded operations enable the generation of remote entanglement 4 times faster than direct transmission at a distance of 100\,km. This simple quantum repeater can be extended by inserting two more nodes, providing a speedup of 82 over 200\,km distance, compared to a system without repeater. The proposed scheme does not rely on assumptions about future technology and thus provides a clear and realistic path toward the experimental demonstration of such a quantum repeater.

\begin{acknowledgments}
This work was supported by the Bundesministerium f\"ur Bildung und Forschung via IKT 2020 (Q.com-Q) and the European Union (Collaborative Project SIQS).
\end{acknowledgments}

\section*{Appendix}
\appendix
\makeatletter
\def\@seccntformat#1{Appendix\ \csname the#1\endcsname:~}
\makeatother

\section{Parameters used for the simulations}
\label{app:Parameters}
We assume a heralding cavity of 400\,\textmu m length. This leaves enough space to put two fiber-based entangling cavities, laterally separated by 200\,\textmu m, between the mirrors, with the fibers having a diameter of 125\,\textmu m. Identical radii of curvature of 500\,\textmu m for the two mirrors of the heralding cavity result in a mode waist of 7.9\,\textmu m at a wavelength of 795\,nm. Using these parameters, the decay rate of $2\pi \times 5.75$\,MHz for the rubidium D$_1$ line and the relative transition strength of $1/4$ for the decay from $5^2P_{1/2} \ket{F'=1;m_F=\pm1}$ to $5^2S_{1/2} \ket{F=2;m_F=\pm1}$, we calculate an atom-cavity coupling strength of $g_\mathrm{h}=2\pi\times 16.3$\,MHz. To implement the scheme with high efficiency, we require mirrors of different transmission: one high reflector and one mirror with higher transmission (called the output coupler) that is the dominant loss channel of the cavity. For the high reflector, we assume a transmission of 10\,ppm and our tests with laser-machined substrates indicate that 20\, ppm parasitic losses per mirror can be achieved for the proposed cavity geometry. In total, this adds up to 50\,ppm losses and a corresponding field decay rate of $\kappa_\mathrm{h}^\mathrm{l} = 2\pi \times 1.5$\,MHz. As a compromise between the directionality of the cavity decay and high cooperativity, we choose a transmission of 400\,ppm for the output coupler, which results in $\kappa_\mathrm{h}^\mathrm{oc} = 2\pi \times 11.9$\,MHz.

An entangling-cavity length of 75\,\textmu m ensures a small mode volume, but also avoids any significant clipping losses of the heralding mode, because even accounting for the curvature of the mirrors, there is still 65\,\textmu m of space between them. We choose asymmetric radii of curvature of 100\,\textmu m for the high reflector and 200\,\textmu m for the output coupler resulting in a mode waist of 4.8\,\textmu m at $\lambda_\mathrm{t}=1476$\,nm. This optimizes the transverse overlap \cite{Joyce1984,Hunger2010} between the cavity mode at the output coupler and the mode of a telecom single-mode fiber (10\,\textmu m mode-field diameter) to 0.96. Due to the asymmetric radii of curvature, the mode radius at the position of an atom trapped in the center of the cavity is 5.3\,\textmu m. The decay rate of the $4^2D_{3/2}$ state to the $5^2P_{1/2}$ state ($5^2P_{3/2}$ state) is $2\pi \times 1.62$\,MHz  ($2\pi \times 0.30$\,MHz) \cite{Safronova2011} and decay from $\ket{F''=1;m_F=0}$ ends up in $\ket{F'=1;m_F=\pm1}$ with $5/12$ probability each. The coupling rate of an atom at the center of the described cavity at this transition is thus $g_\mathrm{t} = 2\pi\times 70$\,MHz. We expect dielectric coatings at telecom wavelengths to have similar performance as coatings for the near-infrared, and thus assume 20\,ppm parasitic losses per mirror for the entangling cavity as well. In combination with 10\,ppm transmission for the high reflector, this results in $\kappa_\mathrm{t}^\mathrm{l} = 2\pi\times 8$\,MHz. We set the transmission of the output coupler to 600\,ppm, resulting in $\kappa_\mathrm{t}^\mathrm{oc} = 2\pi\times 95$\,MHz.

We assume degenerate polarization eigenmodes for the entangling cavity and consider a heralding cavity that only supports a $\pi$-polarized mode, which could effectively be realized by inducing a large frequency splitting between the polarization eigenmodes of the heralding cavity \cite{Uphoff2015}.

For two atoms in one heralding cavity, as required for a full quantum repeater node (Fig.\,\ref{fig:Repeater}), the two atoms cannot both be positioned exactly at the center of the heralding cavity. We propose to place the atoms on the axis of the heralding cavity, $\pm100$\,\textmu m from the center. This leaves enough space for the entangling cavities and results in a small reduction of the coupling to the heralding cavity to $g_\mathrm{h} = 2\pi\times 15.1$\,MHz. This results in only a slight reduction of 1\,\% of the total success probability, which we nevertheless account for in the calculations of the quantum-repeater performance in Sec. \ref{sec:Repeater}.

All parameters mentioned above are consistent with results of ongoing work with optical fiber cavities in our laboratory \cite{Uphoff2015}.

\section{State-dependant probability for herald generation}
\label{app:secondMode}
To see the effect of degenerate polarization eigenmodes, we start with the intermediate entangled state $\ket{\Psi_1}$ in the linear polarization basis for the telecom photon (cf. Eq. (\ref{eq:intermediateStateBasisTrafo}))
\begin{equation}
\ket{\Psi_1} = \frac{1}{\sqrt{2}} \left(\ket{\Psi_\mathrm{1,H}} + \ket{\Psi_\mathrm{1,V}}\right)
\end{equation}
with
\begin{eqnarray}
\ket{\Psi_\mathrm{1,H}} & = &  \frac{1}{\sqrt{2}} \ket{H}_t \left(\ket{-1}_i - \ket{+1}_i\right) \nonumber \\
\ket{\Psi_\mathrm{1,V}} & = & \frac{1}{\sqrt{2}} \ket{V}_t \left(\ket{-1}_i + \ket{+1}_i\right).
\end{eqnarray}
Here, we have assumed $\theta = \pi$ for clarity. In case of $\theta = 0$ the calculation is analogous with similar results.

We assume the polarization eigenmodes of the herald cavity to be $\pi$-polarized and V-polarized. H-polarization is then parallel to the herald cavity axis and would require a longitudinal field which is not supported by the herald cavity. In the worst case, the polarization eigenmodes $\pi$ and V are exactly degenerate, such that the herald cavity enhances decay to states with the same hyperfine quantum number by the same factor. In that case, the final state after emission of a photon into the herald cavity is determined just by the amplitudes of the transitions involved. We assume that free-space decay can be neglected, which is the worst case, because free-space decay is isotropic and cannot lead to a state-dependent probability for herald generation. Under these assumptions, the state $\ket{\Psi_\mathrm{1,H}}$ decays to
\begin{eqnarray}
\ket{\Psi_\mathrm{3,H}}  & = & \frac{1}{\sqrt{\frac{\left|a+a'\right|^2}{2} + \left|b\right|^2 + \left|b'\right|^2 + \frac{\left|c\right|^2+\left|c'\right|^2}{2}}}  \ket{H}_t \nonumber \\
& \times & \left[\ket{\pi}_h\left(b \ket{-1}_f - b' \ket{+1}_f\right)\right. \nonumber \\
& + & \frac{1}{\sqrt{2}} \ket{V}_h\left(c \ket{m_F = - 2} + c' \ket{m_F = +2}\right) \nonumber \\
& + & \frac{1}{\sqrt{2}}\left. \ket{V}_h(a + a') \ket{m_F = 0}\right], 
\end{eqnarray}
where $\ket{m_F=0}$ and $\ket{m_F=\pm2}$ have the same hyperfine quantum number as $\ket{\pm1}_f$. $a$, $b$, and $c$ are the transition amplitudes from $\ket{-1}_i$ to $\ket{m_F=0}$, $\ket{-1}_f$, and $\ket{m_F=-2}$, respectively, and $a'$, $b'$, and $c'$ are the transition amplitudes from $\ket{+1}_i$ to $\ket{m_F=0}$, $\ket{+1}_f$, and $\ket{m_F=+2}$ (cf. Fig.\,\ref{fig:SchemeEntanglementTelecom}).
Similarly, the state $\ket{\Psi_\mathrm{1,V}}$ decays to
\begin{eqnarray}
\ket{\Psi_\mathrm{3,V}}  & = & \frac{1}{\sqrt{\frac{\left|a-a'\right|^2}{2} + \left|b\right|^2 + \left|b'\right|^2 + \frac{\left|c\right|^2+\left|c'\right|^2}{2}}}  \ket{V}_t \nonumber \\
& \times & \left[\ket{\pi}_h\left(b \ket{-1}_f + b' \ket{+1}_f\right)\right. \nonumber \\
& + & \frac{1}{\sqrt{2}}\ket{V}_h\left(c \ket{m_F = - 2} - c' \ket{m_F = +2}\right) \nonumber \\
& + & \frac{1}{\sqrt{2}}\left. \ket{V}_h(a - a') \ket{m_F = 0}\right]. 
\end{eqnarray}

For symmetry reasons, $a'=\pm a$, $b' = \pm b$ and $c'=\pm c$ and the basis states are chosen in such a way that all transition amplitudes are real. Therefore, either $a+a'=0$ and $\ket{\Psi_\mathrm{1,H}}$ does not decay to $\ket{m_F=0}$,  or $a-a'=0$ and $\ket{\Psi_\mathrm{1,V}}$ does not decay to $\ket{m_F=0}$, because of destructive interference. In either case, the probability to emit a $\pi$-polarized photon is different for these two states, because the normalization factor is different. Therefore, selection on events in which a $\pi$-polarized photon was detected does not yield a state with equal amplitudes, which would be required for a maximally entangled state. Instead, a state with less entanglement is created, depending on the value of $|a|$ compared to $|b|$ and $|c|$.

To calculate the fidelity of this state, we set $a'=a$, $b'=b$ and $c'=c$ for simplicity. All other cases are analogous.  If a $\pi$-polarized photon is detected, the normalized final state $\ket{\Psi_3}$ is
\begin{eqnarray}
\ket{\Psi_3} & = &\frac{1}{ \sqrt{\tilde\mathcal{N}}} \bra{\pi}_h \left(\ket{\Psi_\mathrm{3,H}} + \ket{\Psi_\mathrm{3,V}} \right) \nonumber \\
& = & \frac{1}{ \sqrt{\mathcal{N}}} \left[ \sqrt{2b^2+c^2}\ket{H}_t \left(\ket{-1}_f - \ket{+1}_f\right) \right. \nonumber \\
&+& \left.\sqrt{2a^2+2b^2+c^2}\ket{V}_t \left(\ket{-1}_f + \ket{+1}_f\right)\right],   
\end{eqnarray}
with the normalization constants $\tilde\mathcal{N}$ and $\mathcal{N}=4\left(a^2 + 2b^2+c^2\right)$. If $a\neq0$, this is not a maximally entangled state. Its fidelity with the ideal final state $\ket{\Psi_2}$ ($\tilde\theta = \pi$) is
\begin{equation}
F = \left|\braket{\Psi_2|\Psi_3}\right|^2 = \frac{1}{2} + \frac{1}{2} \frac{\sqrt{(2a^2+2b^2+c^2)(2b^2+c^2)}}{a^2+2b^2+c^2}.
\end{equation}

For the transition from $5^2P_{1/2} \ket{F'=1}$ to $5^2S_{1/2} \ket{F=2}$ in $^{87}$Rb the relative transition amplitudes are $a=a'=-1$, $b=b'=\sqrt{3}$, and $c=c'=-\sqrt{6}$, which results in a slight fidelity reduction of $1-F = 0.15\,\%$, even if the polarization eigenmodes of the herald cavity are degenerate. However, for different transitions with a larger $a$ relative to $b$ and $c$, the effect can be much worse. E.g., for the transition from $5^2P_{1/2} \ket{F'=1}$ to $5^2S_{1/2} \ket{F=1}$ in $^{87}$Rb, the transition amplitudes are $a=-a'=-1$, $b=-b'=1$, and $c=c'=0$, which reduces the fidelity by $1-F=2.9\,\%$. As the relative transition amplitudes depend just on the coupling of electron angular momentum, electron spin, and nuclear spin, the effect is the same across different atomic species with the same values for these properties.

\section{Required storage times}
\label{app:RequiredStorageTime}
To estimate the required coherence time of the memories, we study the protocol in more detail. We calculate the time between the first successful creation of entanglement that is still used at the end of the protocol, i.e., is not discarded due to a failed entanglement-swapping attempt, and the end of the protocol. We neglect the possibility of two pairs becoming entangled during the same cycle, which is justified if the entanglement probability $p_\mathrm{e}$ is small. We assume a simple, synchronous temporal operation of the the protocol such that the time interval between two attempts has a fixed length of $L_0/c_\mathrm{f} + \tau$. We assume all memories to decohere equally and it is therefore irrelevant whether the entanglement is transferred to another memory via entanglement swapping.

We first study the protocol that is restarted from the beginning whenever an entanglement-swapping attempt fails. In this case, the expected number $\left<m\right>$ of cycles the entanglement has to be stored is simply the expected number of attempts required to entangle $N-1$ pairs plus one to account for the trial it takes to entangle the memory itself:
\begin{equation}
\left<m\right> = Z_{\mathrm{N}-1}\left(p_\mathrm{e}\right) + 1.
\end{equation}

If the entanglement is kept in case of a failed, unrelated entanglement-swapping attempt, the explicit probability distribution for succeeding in a particular trial is difficult to write down. However, we can extract the maximum number of trials a memory has to store entanglement using the Monte Carlo simulation as in the calculation of the entangled-pair rate. We take $\left<m\right>$ to be the average over $10^6$ runs. Using a bootstrap on the values gathered from these runs, we estimate the 95\,\% confidence interval to be smaller than $\pm 0.2\,\%$ for all values.

\section{Secret fraction}
\label{app:SecretFraction}
To calculate the unconditional secret fraction of a key, which has been generated by a quantum-key-distribution protocol, the density matrix of the imperfect final state has to be known. We assume perfect input states and start with the entangled state after a photonic BSM with contrast $C$. The latter is the probability that the photons interfere and produce the correct entangled state. If they do not interfere, the result of the measurement is a classically correlated state. If the result of the photonic BSM indicates e.g., an entangled state $\ket{\Psi^+} =(\ket{10}+\ket{01})/\sqrt{2}$ between two atoms, the density matrix for the resulting mixed state is
\begin{equation}
\rho_\mathrm{1} = C \ket{\Psi^+} \bra{\Psi^+} + \frac{1}{2} \left(1-C\right) \left( \ket{10} \bra{10} + \ket{01} \bra{01}\right).
\end{equation}

We neglect degradation of the entangled state due to decoherence of the atom. If this is not given in an experimental implementation, calculation of its detrimental effect on the secret fraction requires detailed knowledge of the decoherence mechanisms. Entanglement swapping using an atomic BSM can be performed by applying a CNOT gate followed by a Hadamard gate, measuring the two atoms and performing single-qubit state rotations at the remote target atoms depending on the result of that measurement.
For perfect input states, this procedure is assumed to produce the maximally mixed state $\rho_{1,4}^\mathrm{m} =  \mathbbm{1}/4$ with probability $(1-P)$, such that the fidelity of the entangled state is given by $(1+3P)/4$. Applying this procedure to a tensor product of $\rho_1$ and the same entangled state between two different atoms, yields the density matrix after entanglement swapping
\begin{eqnarray}
\rho_2  & = & P C^2 \ket{\Psi^+} \bra{\Psi^+} + \frac{1}{4} (1-P) \mathbbm{1} \nonumber \\
& + & P C(1-C)\left( \ket{10} \bra{10} + \ket{01} \bra{01} \right) \nonumber \\
& + & \frac{1}{2} (1-C)^2 P \left( \ket{10} \bra{10} + \ket{01} \bra{01} \right).
\end{eqnarray}
Using the identity $\ket{10}\bra{10} + \ket{01}\bra{01} = \ket{\Psi^+}\bra{\Psi^+} + \ket{\Psi^-}\bra{\Psi^-}$, $\rho_{2}$ can be rewritten in the Bell-state basis
\begin{eqnarray}
\rho_{2} & = &\lambda_1 \ket{\Psi^+}\bra{\Psi^+} + \lambda_2 \ket{\Psi^-}\bra{\Psi^-} \nonumber \\
 &+ &\lambda_3 \ket{\Phi^+}\bra{\Phi^+} + \lambda_4 \ket{\Phi^-} \bra{\Phi^-}
\end{eqnarray}
with
\begin{eqnarray}
\lambda_1 & = & \frac{1}{4} (1 + P + 2 PC^2) \qquad \lambda_3  = \frac{1}{4} (1 - P) \nonumber \\
\lambda_2 & = & \frac{1}{4} (1 + P - 2 PC^2) \qquad \lambda_4  =  \frac{1}{4} (1 - P).
\end{eqnarray}
Following the calculation described in Ref. \cite{Scarani2009}, we calculate the secret fraction for entanglement-based quantum key distribution. The error rates $\epsilon_x$, $\epsilon_y$, $\epsilon_z$ for the three bases are
\begin{eqnarray}
\epsilon_x & = & \lambda_2 + \lambda_4 = \frac{1}{2} \left(1 - PC^2\right), \nonumber\\
\epsilon_y & = & \lambda_2 + \lambda_3 = \frac{1}{2} \left(1 - PC^2\right), \nonumber\\
\epsilon_z & = & \lambda_3 + \lambda_4 = \frac{1}{2} \left(1 - P\right),
\end{eqnarray}
and the unconditional secret key fraction is
\begin{eqnarray}
r = 1- h(Q) & - & \epsilon_z h\left(\frac{1 + \left(\epsilon_x - \epsilon_y\right)/\epsilon_z}{2}\right) \nonumber \\
& - & \left(1-\epsilon_z\right) h \left(\frac{1-\left(\epsilon_x + \epsilon_y + \epsilon_z\right)/2}{1-\epsilon_z}\right),
\end{eqnarray}
with the quantum bit error rate $Q=\epsilon_z$ and the binary entropy $h(p) = -p \log_2(p) - (1-p) \log_2(1-p)$.

The error rates for the next swap level, i.e four elementary links, can be calculated by using $\rho_{2}$ instead of $\rho_{1}$ as the initial state and applying the same procedure: 
\begin{eqnarray}
\epsilon_x^{(N=4)} & = & \frac{1}{2} \left(1 - P^3C^4\right), \nonumber \\
\epsilon_y^{(N=4)} & = & \frac{1}{2} \left(1 - P^3C^4\right), \nonumber\\
\epsilon_z^{(N=4)} & = & \frac{1}{2} \left(1 - P^3\right) = Q^{(N=4)}. 
\end{eqnarray}

\end{document}